# Rotator phases in hexadecane emulsion drops revealed by X-ray synchrotron techniques

Diana Cholakova, Desislava Glushkova, Zhulieta Valkova
Sonya Tsibranska-Gyoreva, Krastina Tsvetkova
Slavka Tcholakova, Nikolai Denkov*

*Department of Chemical and Pharmaceutical Engineering*
*Faculty of Chemistry and Pharmacy, Sofia University,*
*1 James Bourchier Avenue, 1164 Sofia, Bulgaria*

*Corresponding author:
Prof. Nikolai Denkov
E-mail: nd@lcpe.uni-sofia.bg
Tel: +359 2 8161639
Fax: +359 2 9625643

# ABSTRACT

Hypothesis

Micrometer sized alkane-in-water emulsion drops, stabilized by appropriate long-chain surfactants, spontaneously break symmetry upon cooling and transform consecutively into series of regular shapes (Denkov et al., *Nature* 2015, *528*, 392). Two mechanisms were proposed to explain this phenomenon of drop "self-shaping". One of these mechanisms assumes that thin layers of plastic rotator phase form at the drop surface around the freezing temperature of the oil. This mechanism has been supported by several indirect experimental findings but direct structural characterization has not been reported so far.

Experiments

We combine small- and wide-angle X-ray scattering (SAXS/WAXS) with optical microscopy and DSC measurements of self-shaping drops in emulsions.

Findings

In the emulsions exhibiting drop self-shaping, the scattering spectra reveal the formation of intermediate, metastable rotator phases in the alkane drops before their crystallization. In addition, shells of rotator phase were observed to form in hexadecane drops, stabilized by $C_{16}EO_{10}$ surfactant. This rotator phase melts at *ca.* 16.6°C which is significantly lower than the melting temperature of crystalline hexadecane, 18°C. The scattering results are in a very good agreement with the complementary optical observations and DSC measurements.

**INTRODUCTION**

Micrometer-sized droplets in still emulsions usually have spherical shape, because this is the shape with the smallest possible surface area and the related smallest interfacial energy. However, recently two groups of researchers revealed in two independent series of studies[1-8] that emulsion droplets stabilized by appropriate long-chain surfactants might change their shape upon cooling prior to the final drop freezing (crystallization). Thus, fluid particles with various non-spherical shapes were formed, including regular polyhedra, hexagonal, tetragonal and triangular prisms, and long thin fibers, see **Figure 1a-d**. This process of drop self-shaping was observed with different classes of organic substances, including *n*-alkanes, alkenes, alcohols and triacylglycerides.[2] For all these chemical compounds, the existence of intermediate rotator phases has been reported in the literature between the fully ordered crystalline and the isotropic liquid phases.[9-15]

The surface area of the self-shaping drops increases between *ca.* 2 and 10 times, in comparison to the surface area of the initially spherical drops. If this increase occurred at non-zero interfacial tension, it would lead to a significant increase in the drop surface energy as well. Therefore, two alternative explanations have been proposed in the literature to explain the drop shape transformations, observed in various emulsion systems.[1,4]

In the first explanation, we suggested that the drop deformation occurs at positive interfacial tension. The related increase of the drop surface energy is compensated by an enthalpy gain, caused by the arrangement of a small fraction of alkane molecules into thin multilayers of plastic rotator phases next to the drop surface.[1-3] Formation of rotator phases (denoted hereafter by R) has been observed experimentally in alkanes with chain lengths between 16 and 50 C-atoms in bulk, in surface layers and/or in various micro- and nanoconfinements.[9-13,16-24]

The molecules in the rotator phases are characterized by limited positional freedom, while being able to rotate and/or oscillate around their long axis.[9-11] Three types of rotator phases were defined in the literature, depending on their stability: thermodynamically *stable* R phases are observed in a certain temperature range on both cooling and heating; thermodynamically *metastable* R phases exist in a certain range only upon cooling, whereas they are not observed upon heating; the least stable, *transient* rotator phases are observed only upon cooling for a very short period of time.[12] Rotator phases have similar structural characteristics to the α-polymorph phase found in acylglycerols with volume per $CH_2$ group ≈ 0.0255 nm$^3$ and cross-sectional area ≈ 0.20 nm$^2$.[13,25] However, the triglyceride polymorphs behave somewhat differently from the rotator phases in alkanes. The polymorphs' thermodynamic stability increases in the order: α < β' < β.[9,26] Upon cooling the molecules can arrange in either of these polymorphs, depending on the specific conditions (*e.g.* on the cooling rate). However, once the β phase is formed, no further structural transformations occur prior to its melting upon heating. If the triglyceride crystallizes first into α phase upon cooling, upon heating this α phase transforms into the more stable β' and/or β phase in the so-called "melt-mediated" exothermic phase transitions. In contrast, when a crystalline alkane is heated, it melts directly into a liquid phase (if a metastable of transient R phase was formed upon cooling) or it passes through one or several stable R phases before melting via endothermic phase transitions. If an alkane, arranged in a metastable or transient R phase is heated, it melts directly without restructuring into more ordered R or C-phase.

The rotator phases observed in bulk odd-numbered alkanes and bulk even-numbered alkanes with chain length longer than 20 C-atoms are thermodynamically metastable or stable, whereas for the bulk shorter even-numbered alkanes (incl. hexadecane) only transient R phases were reported.[12,13] Depending on the specific arrangement of the molecules in the lamellar rotator phases, five different structures were revealed which are denoted in the literature by Roman numbers in subscript, $R_I$ to $R_V$.[10,11]

For the bulk hexadecane, $C_{16}$, a transient R phase was detected experimentally only in 9 out of 85 cooling cycles with maxima at scattering vector $q$ = 2.86 nm$^{-1}$.[12] The stable crystalline triclinic phase (denoted as C for crystal) appeared at $q$ = 3.05 nm$^{-1}$. These q-vectors correspond to lamella thicknesses of 2.2 nm for the R phase and 2.06 nm for the C phase. However, in crystalline phase the molecules are tilted at about 19.4°, therefore the lamellar thickness in the R phase indicates that the molecules are probably in the untilted $R_I$ rotator phase.[12] Similar length of the straight hexadecane molecule, 2.23 nm was reported in Ref. [9].

Numerous studies have shown that the confinement effects and/or the alkane mixing increase significantly the stability of the intermediate rotator phases.[13,18-24] For example, the formation of rotator phase $R_I$ was observed for $C_{13}$ + $C_{14}$ mixture embedded in 8 nm pores of controlled porous glass[27], whereas both $R_I$ and $R_{II}$ rotator phases were observed for $C_{16}$ + $C_{18}$ alkane mixture[28].

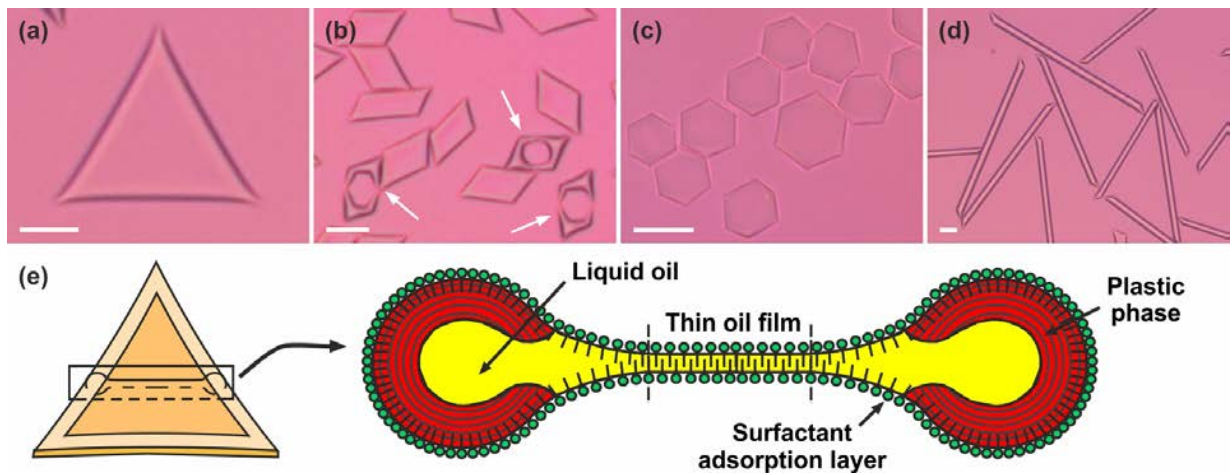

**Figure 1. Microscopy images of non-spherical hexadecane drops (a-d) and schematics of the rotator phase mechanism (e).** (a) Triangular platelet; (b) Tetragonal platelets; three of these platelets (shown by white arrows) have a donut-like shape, i.e. they have a hole in their center while still being in a fluid state; (c) Hexagonal platelets; (d) Rod-like fluid drops. All images are of hexadecane drops dispersed in 1.5 wt. % $C_{16}SorbEO_{20}$ surfactant system. Scale bars, 20 μm. (e) Schematic presentation of a triangular platelet and its cross-section. See the main text for additional explanations.

The structural studies of rotator phases in emulsion drops are still very limited.[29-31] Shinohara, Ueno and co-authors studied the crystallization of $C_{16}$ drops, dispersed in four surfactant solutions – Tween 20, 40, 60 and 80. They showed that upon 2°C/min cooling, the drops stabilized by Tween 40 ($C_{16}SorbEO_{20}$) passed through intermediate rotator phase, whereas no rotator phase formation was detected for Tween 80 ($C_{18:1}SorbEO_{20}$) which contains a double bond in the surfactant tail. In these studies no optical microscopy was used to clarify whether drop shape deformations occurred. Therefore, it remained unclear why R phase was observed in this particular system and whether this intermediate phase was related to any drop shape changes.

In our experiments,[1,2] non-spherical drops were observed only in the emulsions stabilized by surfactants with sufficiently long tails that are up to 3 C-atoms shorter, similar or longer than the alkane chains in the emulsion drops and with not-too-voluminous headgroups. Therefore, we proposed that the formation of rotator phase is caused by freezing of the surfactant adsorption monolayer which then serves as a template for ordering of the neighboring alkane molecules into multilayers of plastic rotator phases. These multilayers possess sufficient mechanical strength to counteract the capillary pressure caused by the curved interface at non-zero interfacial tension, thus allowing for the formation of non-spherical fluid particles. Schematic presentation of the rotator phase mechanism[1-3] driven by the interfacial crystallization process[32,33] is shown in **Figure 1e**. This mechanism was implemented in a detailed theoretical model[34,35] which predicts the observed sequence of drop-shape transformations from first principles (energy minimization).

An alternative explanation for the observed drop shape deformations was proposed by Guttman and co-authors[4,5,7] who also worked with hexadecane-in-water emulsions, however, stabilized by a different surfactant – the cationic octadecyl trimethylammonium bromide ($C_{18}$TAB). These authors suggested that the observed deformations are caused by the freezing of the surfactant adsorption monolayer only and that the emulsion cooling leads to ultra-low and even to "transiently negative" interfacial tension which allows for an increase of the interfacial area without energy penalty.[4,5,7] In a more recent theoretical study,[36] this group suggested that the drop shape transformations could be explained by considering the spontaneous curvature of the surfactant monolayer, which is again possible only if the interfacial tension is very low, viz. below *ca.* 0.1 mN/m.

The systems studied by Denkov and co-authors were classified in four groups depending on the temperature, $T_d$, at which the drop shape deformations begin. The systems for which spontaneous symmetry break of the drop shape is observed upon cooling and the drops evolve along the main stages of an universal shape sequence belong to *Groups A, B* or *C*. In contrast, the emulsions in which the drops freeze in spherical or spheroidal shape belong to *Group D*.[2] For the emulsions in *Group A*: $T_d > T_m$, for *Group B*: $T_d \approx T_m$ and for *Group C*: $T_d < T_m$, where $T_m$ is the melting temperature of the oil. The interfacial tensions of several representative emulsions from each group were measured as a function of temperature.[3] The results showed interfacial tensions in the range between *ca.* 2-3 mN/m for the systems in *Group A* and 6-8 mN/m for the systems in *Groups B* and *C*, at the temperatures corresponding to the onset of drop deformations. These values are much higher than those that could explain the drop shape transformations by the formation of a frozen monolayer only. Thus, we concluded that thicker surface multilayers are needed to deform the drop surface against the capillary pressure for non-spherical drops to be formed in the emulsions studied by our group.

Following this reasoning, in Ref. [8] we performed differential scanning calorimetry (DSC) measurements of the enthalpy gain during the formation of the non-spherical drops in several emulsion systems. We found that the observed enthalpy gain for $C_{16}$ drops dispersed in $C_{16}EO_{10}$ surfactant (emulsion falling in *Group A*) corresponds to the formation of 2 to 80 ordered layers next to the drop surface, depending on the specific assumption for the distribution of the rotator phase over the drop surface. For the system of 9.5 μm $C_{16}/C_{16}SorbEO_{20}$ (*Group B*) forming tetragonal platelets, thicknesses of *ca.* 40 to 70 nm were calculated, corresponding to 24 ± 6 layers of ordered molecules, and for $C_{16}/C_{16}EO_{20}$ in which thin fibers were observed upon cooling, the thickness of the rotator phase was estimated to be around 8 nm, i.e. 3 ± 1 layers of molecules.

Until now, no direct structural characterization has been attempted for the emulsions studied by us to verify the rotator phase mechanism. Therefore, in the current study we perform time resolved SAXS/WAXS measurements with synchrotron radiation source, varying the temperature in the range in which the drop shape deformations are observed, to obtain direct information about the molecular arrangement inside the deforming alkane drops. Several emulsion systems from the different groups (*A* to *D*) were studied. The obtained X-ray spectra are complemented with optical microscopy observations to make a link between the drop shapes and the possible formation of rotator phases in them.

The obtained results reveal that rotator phases form only in the emulsions in which drop shape transformations are observed. The experiments allowed us to characterize these rotator phases and to determine some of their properties. In one of the studied emulsions we observed thick shells of rotator phase to form next to the drop surface. In another emulsion, stabilized by a different surfactant, complete plastification of the particle volume (*viz.* liquid to rotator phase transition) was observed. Two different in structure, metastable rotator phases were detected in these emulsions and their melting temperatures were determined.

**MATHERIALS AND METHODS**

**MATERIALS**

Linear *n*-hexadecane ($C_{16}H_{34}$), denoted in text as $C_{16}$, was used as disperse phase in the tested emulsions. The alkane was purchased from Sigma-Aldrich (purity 99%) and had melting temperature of $T_m = 18°C$. Before we used it for emulsions preparation, we purified it from possible surface-active contaminations by passing it three times through a glass column filled with Florisil adsorbent.

Four nonionic water-soluble surfactants were used for emulsion stabilization: polyoxyethylenesorbitan monolaurate, $C_{12}SorbEO_{20}$, trade name Tween 20; polyoxyethylenesorbitan monopalmitate, $C_{16}SorbEO_{20}$ (Tween 40) and polyoxyethylene (10) cetyl ether, $C_{16}EO_{10}$ (Brij C10). These surfactants were purchased from Sigma-Aldrich and were used as received, without further purification. In one series of experiments we used also Lutensol AT50, product of BASF, which contains mixture of $C_{16}$ and $C_{18}$ alcohol ethoxylates with average number of 50 ethoxy units (denoted hereafter as $C_{16-18}EO_{50}$). Note that all these surfactants contain mixtures of molecules and these mixtures may vary from one batch to another. In the current study, Tween 40 surfactant was from a new batch, different from that used in our previous studies, for example in Ref. [8]. Therefore, the beginning of the drop shape deformations here was observed at slightly higher temperatures than before (while the observed drop shape transformations were the same), most probably because the fraction of the longer-chains was slightly higher in the new surfactant batch.

All aqueous solutions were prepared with deionized water purified by Elix 3 module (Millipore) which had resistivity > 18 MΩ·cm. The surfactant concentration was 1.5 wt % for $C_{12}SorbEO_{20}$, $C_{16}SorbEO_{20}$ and $C_{16-18}EO_{50}$, and 0.75 wt. % for $C_{16}EO_{10}$. These concentrations are well above the critical micellar concentrations (CMC) of these surfactants.

**METHODS**

*Emulsion preparation.*

The studied emulsions were prepared using membrane emulsification technique which allows production of relatively monodisperse in size droplets.[37,38] The emulsification procedure used was similar to the one used in our previous studies, see Refs. [2,8] for detailed explanations. In the current study, we used porous membranes with pore diameters of 4 μm, 5 μm and 10 μm to prepare drops which have diameter about three times bigger than the membrane pore size.

*SAXS/WAXS measurements*

The simultaneous small- and wide-angle X-ray scattering measurements were performed in Elettra Sincrotrone, Trieste, Italy at the Austrian SAXS beamline. The experiments were performed at fixed cooling and heating rates of 0.5°C/min. The working energy was 8 keV (λ ≈ 1.55 Å). Two separate detectors were used: 2D Pilatus3 1M detector for collection of the SAXS images (detector to sample distance of 123 cm) and 2D Pilatus 100k for WAXS signal. The obtained signal covers scattering vectors, $q = 2\pi/d$, between 0.1 and 5.7 nm$^{-1}$ and from 8.4 to 18.4

nm$^{-1}$, respectively. The SAXS and WAXS scans were obtained with 14.9 s exposure time and single scattering profile was measured at every 15 s. The samples were placed in cylindrical borosilicate glass capillaries with 1.5 mm diameter, 80 mm length and 0.01 mm wall thickness, which were placed in aluminum thermostatic chamber connected to a thermostat during the measurements. The temperature was measured with a calibrated thermocouple probe inserted in another identical borosilicate glass capillary, filled with water and sealed with wax, which was inserted in a neighboring orifice in the thermostatic chamber. The latter capillary was identical in dimensions to the one used for the actual SAXS measurements.

The SAXS/WAXS spectra are presented in the paper as originally obtained from the measurements, except that some of the curves are shifted with respect to their intensities (y-axis) for clearer visualization. The background signal from the surfactant solution is a signal with almost constant intensity and, therefore, it does not influence the observed pattern in the relevant *q*-range. For this reason, we have not subtracted it from the original data, as it would change only the absolute intensity values which are not used for the data interpretation. SAXS spectra of the surfactant solutions are present in **Supplementary Figure S1**.

*Optical observation of drop-shape changes in a capillary*

The optical observations were performed with AxioImager.M2m microscope (Zeiss, Germany) in transmitted, cross-polarized white light. A λ-compensator plate was placed after the studied specimen and before the polarizer at 45° angle with respect to both the analyser and polarizer to achieve conditions under which the fluid objects have typical magenta color, whereas the frozen birefringent objects have intense colors.[39,40] Long-focus objectives with magnifications ×20 and ×50 were used for the observations.

The experiments were performed using a specimen placed in glass capillary with rectangular cross-section (50 mm length, 2 mm width, 0.2 mm height) which was enclosed within a custom-made cooling chamber connected to cryo-thermostat (JULABO CF30, Cryo-Compact Circulator). We used calibrated thermo-couple probe with an accuracy of ± 0.2°C for measurements of the temperatures during the experiments. More details about the exact temperature calibration procedure ensuring that the measured values are maximally precise are available in Ref. [8].

Microscopy pictures in white transmitted light were used for determination of the drop-size distribution in the original emulsions, prior to the cooling and heating experiments, denoted in text as $d_{ini}$. We measured the diameter of more than 6 000 individual droplets using the Image Analysis Module of Axio Vision Software for each sample. The Sauter mean diameter $d_{32}$ was calculated, $d_{32} = \sum_i N_i d_i^3 / \sum_i N_i d_i^2$, where $N_i$ is the number of drops with diameter $d_i$ and will be used hereafter as $d_{ini}$.

### DSC experiments

Discovery DSC 250 equipment (TA Instruments, USA) was used for the differential scanning calorimetry measurements. Fixed cooling and heating rate of 0.5 °C/min was used. A more detailed explanation of the DSC procedure is available in Ref. [8].

**RESULTS AND DISCUSSION**

In this section we present our results from the SAXS measurements and connect them to the drop-shape transformations observed by optical microscopy and to the enthalpy effects measured by DSC.

The scattering profile observed with bulk hexadecane is presented in **Supplementary Figure S2**. The rotator phase in bulk hexadecane is thermodynamically unstable and has been observed under specific conditions for a short period of time only.[12] As seen from the curves in **Supplementary Figure S2**, we did not detect rotator phase formation in the bulk hexadecane with the protocols used in our experiments. Therefore, the rotator phase observed to form in some of the emulsions can be attributed only to arrangement of the alkane molecules inside the micrometer drops. In addition, the observed supercooling for nucleation in bulk hexadecane, by ≈ 1.5°C below $T_m$ = 18°C, is in excellent agreement with the freezing temperature for bulk hexadecane, as reported by Sirota and Herhold, $T_f ≈ 16°C$.[12]

### $C_{16}$ drops in $C_{12}SorbEO_{20}$ surfactant

First, we present the results obtained in experiments with 12.5 ± 0.1 μm drops of $C_{16}$ dispersed in $C_{12}SorbEO_{20}$ solution. This system falls in *Group D*, *i.e.* this is a system in which no drop shape changes are observed prior to the final drop freezing.[2] The reason for the absence of drop deformations in this system is that the surfactant molecules have too short hydrophobic tails (4 C-atoms shorter than the dispersed alkane) and do not freeze in the adsorption layer prior to the freezing of the alkane in the drop interior. Thus, no template for formation of plastic rotator phase is created before the alkane freezes into a crystal phase.

**Figure 2a** presents the SAXS profile measured for this system at a constant cooling rate of 0.5°C/min, from 20°C to 5°C. The appearance of the peak for triclinic crystal structure, $q \approx 3.05$ nm$^{-1}$, begins at 10.9 ± 0.4°C. The intensity of the initial observed peak is ≈ 1.1% from the intensity of the fully crystallized drops at 5°C, as determined from 3 independent experiments. This peak gradually increases its intensity to about 19 ± 2% at 9°C, 56 ± 2% at 7°C and 72 ± 2% at 6°C when compared to the peak observed at 5°C. The small standard deviation calculated from the three independent experiments shows that the results are very reproducible.

Peak with maximum at $q \approx 2.8$ nm$^{-1}$, corresponding to rotator phase, was not detected upon cooling or upon heating in any of the experiments with $C_{16}/C_{12}SorbEO_{20}$ emulsion in the entire range of temperatures studied, see the inset in **Figure 2a**.

The observed SAXS profile is in complete agreement with the results obtained from the optical observations. As seen from **Figure 2b**, the first drops crystallize at temperature around 10-11°C, while the main fraction of the drops freeze in the range between 8 and 5°C.

Therefore, we conclude that plastic rotator phase is not forming in the system $C_{16}/C_{12}SorbEO_{20}$ and, respectively, the drops crystallize in spherical or close to spherical shapes without passing through any non-spherical shapes.

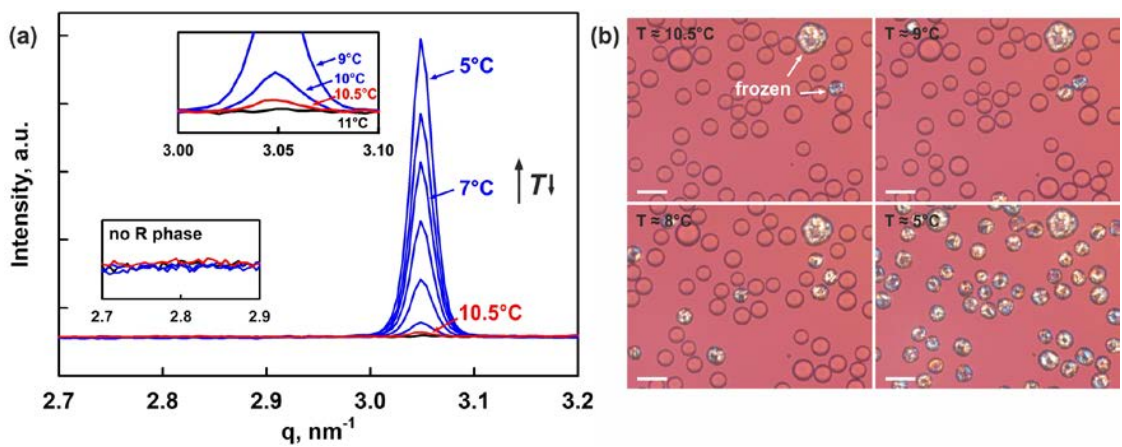

**Figure 2. SAXS profile and optical microscopy images of 12.5 µm $C_{16}$ drops in $C_{12}SorbEO_{20}$ solution.** (a) SAXS profile. The black line shows the signal measured at 11°C, the red line shows the signal at 10.5°C, and the series of blue lines shows the signal measured between 10°C and 5°C, at 1°C temperature interval. Insets: zoom-in of the signal in the $q$-range between 2.7 nm$^{-1}$ and 2.9 nm$^{-1}$ where the peak of the rotator phase is expected to appear and of the $q$-range between 3 and 3.1 nm$^{-1}$ where the peak of the crystal phase appears. **(b)** Microscopy images obtained at different temperatures, illustrating the freezing of the dispersed drops in spheroidal shape. Scale bars, 20 μm.

To verify our conclusion about the systems in *Group D*, we performed additional experiments with another surfactant-oil combination from this group, namely, hexadecane drops dispersed in the ethoxylated nonionic surfactant with 16-18 C-atoms in the tail and 50 ethylene oxide units in the hydrophilic head, $C_{16}/C_{16-18}EO_{50}$. In this case, the surfactant tail is with appropriate length, but the very voluminous head group suppresses the formation of dense surfactant adsorption layer and, as a result, no significant drop shape deformations were observed in this system as well.[2]

In the SAXS signal obtained from this emulsion, we observed a very weak but still detectable peak (the greatest area of the peak was ≈ 0.8% from the maximal area of the C peak) for rotator phase at $q ≈ 2.82$ nm$^{-1}$ which appeared when the drops began to crystallize around 11.2°C, together with the peak for the crystalline phase, $q ≈ 3.05$ nm$^{-1}$, see **Figure 3a**. The crystalline peak increased its intensity with the decrease of temperature, whereas the rotator phase peak remained visible with very low intensity until the crystallization of all drops finished at $T ≈ 5°C$ and then this peak disappeared. The crystalline peak achieved its maximal intensity at this temperature, showing that no further crystallization occurred in the sample.

To understand the reason for the formation of this very small but detectable peak of the rotator phase we repeated our optical microscopy observations using a video-camera with higher frame rate to check whether any quick shape deformations occurred in these emulsions prior to drop freezing. Indeed, we observed that just before freezing, some of the fluid drops became slightly corrugated and immediately afterwards froze, see **Figure 3b**. This result shows that the freezing of the surfactant adsorption layer and the freezing of the alkane molecules occur at the same temperature and, therefore, the drops have no time to change their shape.

To check whether the formation of non-spherical drops was kinetically suppressed and whether we could induce the formation of thicker layers of plastic rotator phase prior to drop freezing, we performed additional experiments at 5 times lower cooling rate of 0.1°C/min. However, no significant difference in the drop behavior was observed and the drops froze in a spheroidal shape.

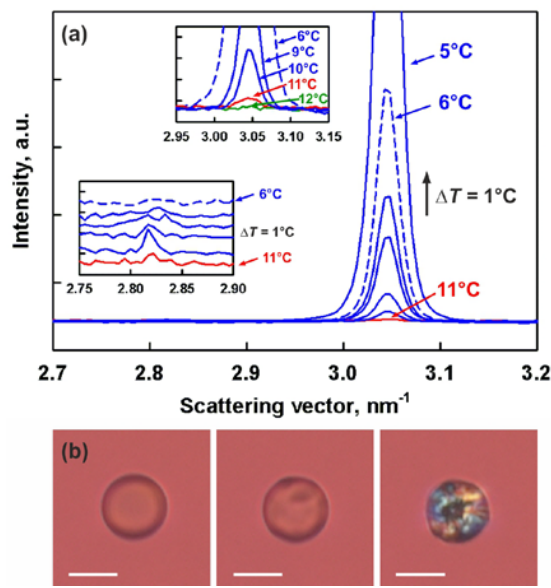

**Figure 3. SAXS profile and microscopy images of 13 μm $C_{16}$ drops in emulsion stabilized by $C_{16-18}EO_{50}$ surfactant.** (a) SAXS profile obtained upon cooling. One main peak showing the formation of triclinic crystalline phase is observed at $q \approx 3.05$ nm$^{-1}$. Insets: zoom-in of the signal in the $q$-range between 2.7 nm$^{-1}$ and 2.9 nm$^{-1}$ where a very small peak of rotator phase appears when the crystallization begins and disappears when it ends; the second inset shows the $q$-range between 3 and 3.1 nm$^{-1}$ at the onset of drop crystallization. (b) Sequence of microscopy images of hexadecane drop before, in the moment prior to, and after crystallization – some surface corrugations are observed just before the crystallization. Scale bars, 10 μm.

This surfactant ($C_{16-18}EO_{50}$) is of technical grade and contains a mixture of molecules with 16 and 18 carbon atoms and with various numbers of EO units, attached during its manufacturing.[41,42] The number 50 given by the producer indicates the average value of the EO units per molecule. The obtained results indicate that the observed R phase is most probably caused by the presence of surfactant molecules with smaller number of EO units in this surfactant mixture which are able to compact together and freeze just before the oil freezing.

Summarizing, two main types of emulsion from *Group D*, i.e. systems in which significant drop shape deformations prior to the freezing are not observed, were carefully studied: (1) Oil-surfactant combination for which the surfactant tail is by 4 C-atoms shorter than the alkane length. For this system, no rotator phase is detected in the SAXS measurement and, respectively, no drop shape deformations were observed upon cooling; (2) System in which the surfactant head group is too voluminous and suppresses the formation of dense adsorption layer, while the surfactant tail is comparable in length to the alkane molecules. In this case, we showed that very small rotator phase peak can be detected which can induce only small corrugations on the drop surface, just before drop freezing. However, this small quantity of rotator phase is unable to induce the formation of regular in shape, non-spherical drops, even when very low cooling rate of 0.1 °C/min is used.

### $C_{16}$ drops in $C_{16}SorbEO_{20}$ surfactant

Next, we present the results obtained upon 0.5°C/min cooling of emulsions of $C_{16}$ drops with $d_{ini} \approx 12.1 \pm 0.1$ μm, stabilized by $C_{16}SorbEO_{20}$ surfactant, see **Figure 4** and **Supplementary Figure S3**. The fraction of the rotator phase formed in this system is expected to be the highest from the results obtained in our previous DSC experiments.[8] Initially only one peak appears at $q \approx 2.82$ nm$^{-1}$ in the SAXS spectra, showing the formation of a rotator phase at $T \approx 16.1 \pm 0.2°C$ (data from 8 independent experiments). Upon further cooling, a very small crystalline peak appears at $q \approx 3.05$ nm$^{-1}$ at $T \approx 15.2 \pm 0.6°C$. The rotator phase peak remains with higher intensity than the crystalline phase peak down to *ca.* $T \approx 12.5 \pm 1°C$ where the two peaks become comparable. The intensity of the C peak increases significantly in the temperature range of 12 to 10°C, but the complete disappearance of the rotator phase peak is observed at much lower temperature, $8.4 \pm 0.1°C$. The crystalline peak increases slightly its intensity upon further cooling down to 7°C. No rotator phase is detected upon heating of this sample. When heated, the C phase peak begins to decrease around 17.6°C and then disappears completely upon further temperature increase.

To interpret precisely the nature of all peaks we performed optical microscopy observations with concentrated emulsion samples, similar to those used in the synchrotron experiments, see **Figure 4b**. Data from 8 independent experiments showed that the drop shape deformations first appeared at $T \approx 16.4 \pm 0.5°C$ and all drops obtained polyhedral shape down to $T \approx 15.8 \pm 0.6°C$. Interestingly, colors indicating the formation of highly ordered phase in the entire drop volume were observed in a very small fraction of the drops at a relatively high temperature of $T \approx 15.6 \pm 0.2°C$, see **Figure 5b**. We counted the number of the drops undergoing such phase transition to be 26 only, out of more than 11 500 observed individual drops, which corresponds to around 0.2 %. The other droplets changed their shape following the previously reported sequence of shape transformations from spheres via polyhedrons and hexagonal platelets into trigonal and (predominantly) tetragonal platelets. A fraction of the fluid platelets was observed to freeze at 11-12°C, whereas the others evolved finally into rod-like particles and froze at 7 ± 1°C, see **Figure 4b**.

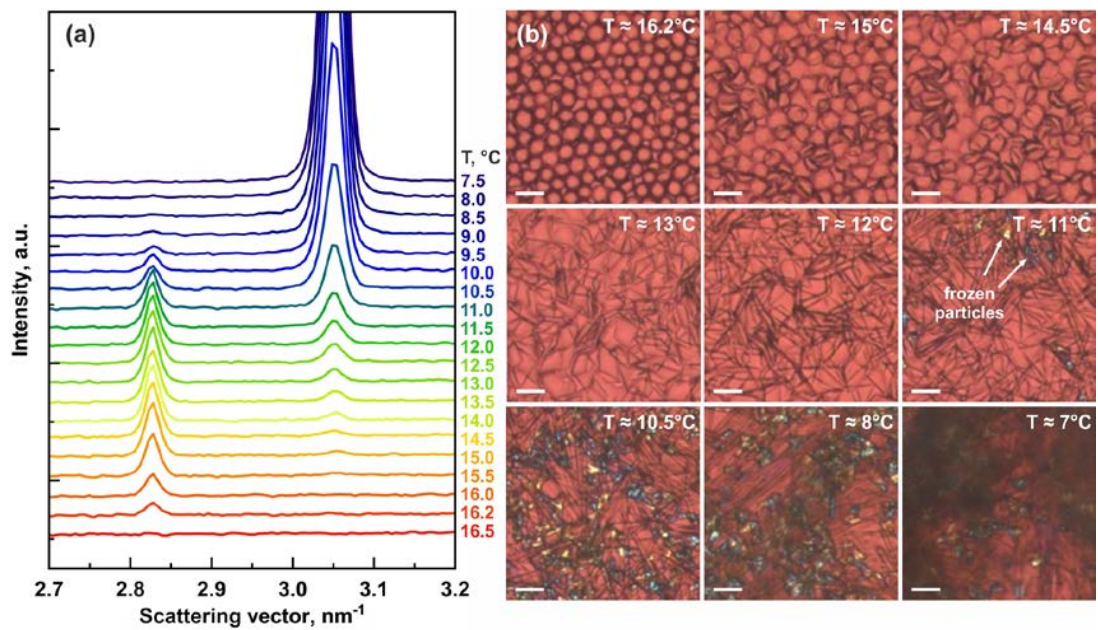

**Figure 4.** **SAXS profile and microscopy images observed with 12.1 μm $C_{16}$ drops in $C_{16}SorbEO_{20}$ solution.** **(a)** SAXS profile obtained upon cooling at 0.5°C/min. The numbers on the right indicate the temperature at which the respective scan was taken. **(b)** Microscopy images obtained at different temperatures, demonstrating the deformations and freezing of the dispersed drops. Scale bars, 20 μm. For WAXS spectrum, obtained from this sample and for enlarged view of the initial peaks, see **Supplementary Figure S3**.

The following three hypotheses can be formulated by comparing the results of the two experiments: (1) The R phase peak is caused by the formation of non-spherical fluid particles, whereas the small C peak which appears at relatively high temperatures is caused by the freezing of this very limited number of particles; (2) The R phase peak is caused by the plastification, *viz.* by the liquid-to-rotator phase transition, of the limited number of particles obtaining bright colors at high temperature, whereas the C phase peak at high temperatures shows their transition from R to C phase, and (3) A combination of explanations (1) and (2). To test these hypotheses, additional experiments were performed with the following temperature protocol: the samples were cooled from 20°C allowing the particles to obtain non-spherical shapes and then the cooling was stopped before the crystallization of the main fraction of the drops had occured. Afterwards, the samples were slowly heated until disappearance of all peaks in the SAXS/WAXS spectra or

*Paper accepted for publication in Journal of Colloid and Interface Science, doi: 10.1016/j.jcis.2021.06.122*

until the formation of spherical liquid drops in the microscopy experiments. Illustrative results from these experiments are presented in **Figures 5** and **6**.

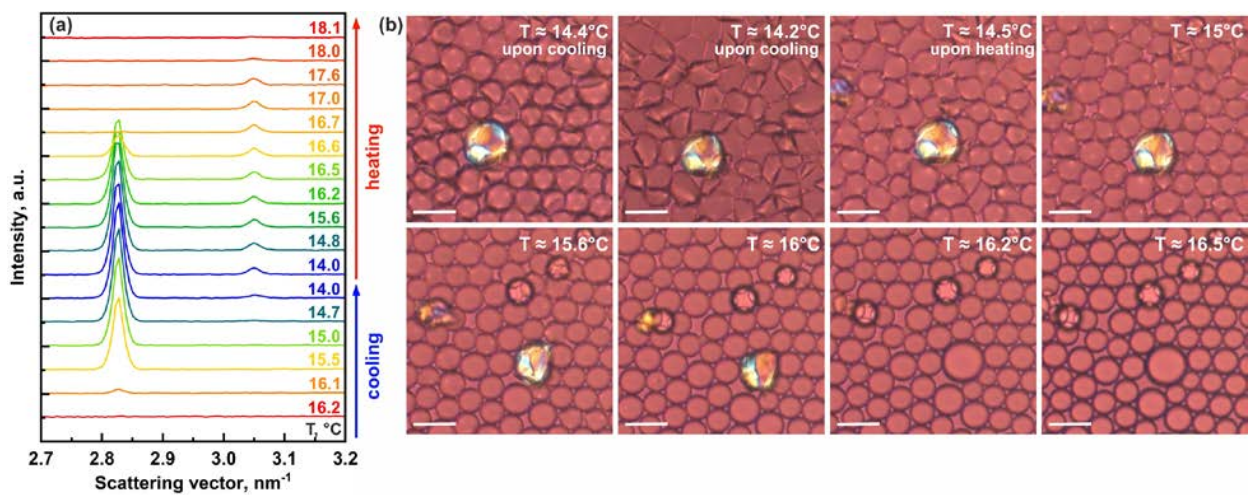

**<u>Figure 5.</u>** **SAXS profile and microscopy images observed with 12.1 μm $C_{16}$ drops in $C_{16}SorbEO_{20}$ solution without complete freezing of the droplets. a.** SAXS profile obtained upon cooling at 0.5°C/min to 14°C. Afterwards the sample is heated at 0.5°C/min up to 20°C. The R phase peak disappears around 16.7°C, whereas the small C peak disappears when the melting temperature of the bulk hexadecane is reached. The numbers on the right indicate the temperature at which the scans have been taken. **b** Microscopy images obtained upon cooling to 14°C and subsequent heating up to 16.5°C. Upon cooling, very limited fraction (< 0.2%) of the particles freezes and acquires bright colors, while the rest of the drops deform and remain fluid. If the sample is not cooled down to induce further drop freezing, the deformations in the fluid drops disappear along with the melting of these colored particles up to 16.5°C. Scale bars, 20 μm.

As seen from the presented cooling curves in **Figure 5**, the R phase peak appears around 16.1°C, whereas the C phase peak appears at lower temperature of around 14.7°C. Note that the maximal peak area of C-phase peak is about 1.9% only from the area of the final crystalline peak, obtained upon complete freezing of all drops in the sample. The small C peak slightly increases upon cooling to 14°C. Upon heating, the crystalline peak does not change until the melting temperature of the bulk hexadecane is reached, 18°C, whereas the rotator phase peak gradually decreases its intensity and completely disappears at 16.7°C. Similar results were obtained in all experiments of this type. The average temperature at which the R peak disappeared completely was 16.7 ± 0.3°C (data from 4 independent experiments). The microscopy observations showed that the initial particles which had obtained bright colors, melted back into fluid droplets up to this particular temperature. In addition, all drop deformations disappeared also up to this temperature. Therefore, these "colored" particles underwent a plastification process and their entire volume was structured in a plastic rotator phase, causing the appearance of the bright colors. The melting of these particles, composed of rotator phase, occurred at a temperature which was by ≈ 1.5°C lower than that of the crystal phase. The small crystalline peak which disappeared at 18°C was most probably caused by the melting of a very limited fraction of crystalline particles which had transformed from R to C phase upon cooling (nucleation is a stochastic process).

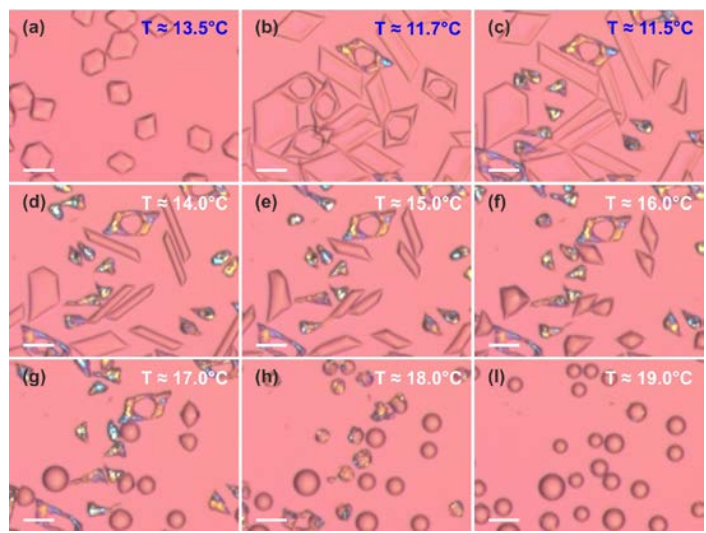

**Figure 6.** **Microscopy images of $C_{16}$ drops dispersed in $C_{16}SorbEO_{20}$ solution.** (a-c) Drops observed upon cooling. Note that 5 of the drops in (b) have holes in their centers while still being in a fluid state. The single particle with bright colors is frozen. The cooling is stopped at 11.5°C and then heating is started. (d-i) Drops observed at different temperatures upon heating. The deformations of the fluid drops disappear at 16-17°C, whereas the frozen particles melt into spherical drops when the melting temperature of bulk hexadecane is reached, 18°C. Scale bars, 20 μm.

To test further whether the shape deformations disappear at temperatures significantly lower than the hexadecane melting temperature, if the drops have evolved further down in the evolutionary sequence, we performed additional microscopy experiment by cooling the sample down to 11.5°C and then heating it back up to room temperature, see **Figure 6**. In this case, part of the drops froze, whereas the other drops remained in a fluid state. Note that cooling to lower temperature allows additional time for plastic rotator phase formation on the deforming drop surface. As seen from these images, although the shape deformations were much more pronounced, most of the drops obtained spheroidal shape in the range between 16°C and 17°C. In contrast, the particles which had frozen at low temperature remained frozen until the crystal melting temperature of 18°C was reached.

From all these results we conclude that the peak of the R phase appearing in the SAXS spectra at *ca.* 16.2°C is caused by the formation of both deformed particles (surface rotator phase) and bulk plasticized particles (particles composed entirely of R phase) which melt at temperature which is by ≈ 1.5°C lower than the melting temperature of the bulk crystalline hexadecane. Note that no R phase is observed upon heating, after the R→C phase transition has been completed upon cooling. The latter result evidence that the observed R phase is metastable. In other words, it forms upon cooling and exists for a long time in a certain temperature interval before transforming into the more stable crystalline phase, but it does not form upon heating from the crystalline phase. To the best of our knowledge, such metastable R phase has never been reported in hexadecane emulsion droplets so far and it is certainly related to the specific surfactant used in these experiments.

### *$C_{16}$ drops in $C_{16}EO_{10}$ surfactant*

To complete our study, we tested also $C_{16}$ drops with two different initial diameters, 33.5 μm and 15 μm, dispersed in $C_{16}EO_{10}$ solution. In this emulsion, the drop shape deformations begin at $T_d \approx$ 21-22.5°C > $T_m$ = 18°C.[8] In our previous study we showed that around the deformation temperature, $T_d$, small but reproducible DSC peak was observed in the thermograms (Figure 7 in Ref. [8]). Upon cooling, the drops in this system transform into polyhedrons with irregular shape and corrugated surface which break spontaneously into smaller droplets several times before their final freezing at *ca.* 15 ± 0.5 °C, see Supplementary Movie 1. After each breakage event, several smaller drops are formed but the main fraction of the oil remains in the initial "mother" drop.

The DSC measurements with this system showed that the heat curve remains above its baseline value after the beginning of the deformations, reflecting the continuous drop shape transformations down to 15.1°C ± 0.1°C where strong exothermic peak appeared, **Figure 7a**. After this large peak, several additional exothermic peaks appeared in the range between 14.8°C and 9°C. The final broad peak between 9°C and 11.5°C is due to the freezing of smaller drops produced in the breakage events, as described above.[8] The origin of the peaks observed between 11.5°C and 14°C remained somewhat unclear before the SAXS measurements were performed for the current study.

The measured SAXS profiles for 33.5 μm are shown in **Figure 7b,d** and the WAXS profiles – in the **Supplementary Figure S4**. No peaks were detected at temperatures 21-22°C where the drop shape deformations start, showing the insufficient sensitivity of the equipment to detect a limited number of ordered surface layers. The first deviation from the baseline value of the measured signal is observed at 15.5 ± 0.2°C (averaged from 10 independent experiments) with a peak at $q \approx 2.82$ nm$^{-1}$, representing the rotator phase formation, see **Figure 7b** and **Supplementary Figure S4a**. A very small peak for triclinic crystalline phase appears at $q \approx 3.05$ nm$^{-1}$ after 1-2 minutes, when the temperature was decreased down to 15.1 ± 0.3°C.

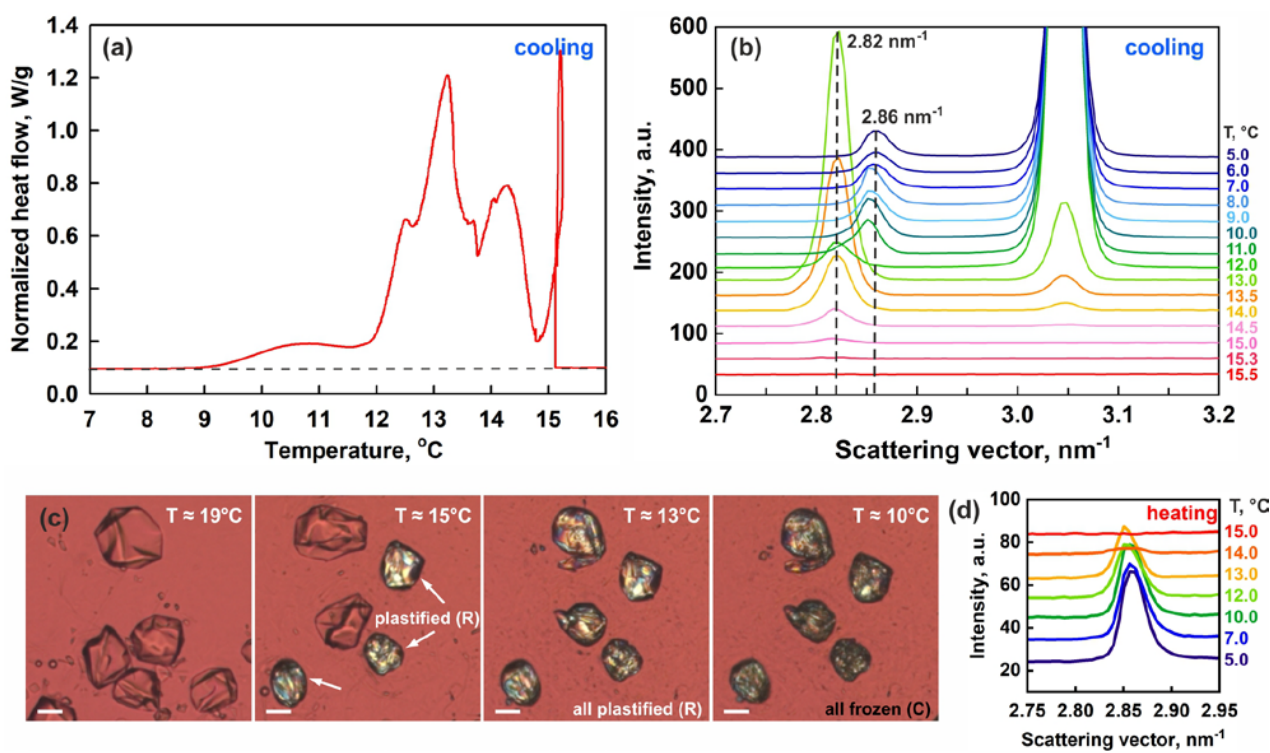

**Figure 7.** Results obtained with 33.5 μm $C_{16}$ drops dispersed in $C_{16}EO_{10}$ solution. (a) DSC, (b) SAXS and (c) microscopy images, obtained at 0.5°C/min cooling. (d) SAXS signal in the $q$-range of 2.75 to 2.95 nm$^{-1}$ obtained during 0.5°C/min heating of the same sample as shown in (b). For explanations of the observed processes, see the main text. Scale bars on (c) = 20 μm. See also **Supplementary Figure S4** for enlarged view of the beginning of the plastification process and WAXS curves.

We note that the intensity of the R peak in these samples is significantly higher than the intensity of the R peak observed in the $C_{16}/C_{16}SorbEO_{20}$ samples. While the maximal peak area of R phase in $C_{16}/C_{16}SorbEO_{20}$ was about 8.5 ± 4 % of the maximal C peak area for the same samples, for $C_{16}/C_{16}EO_{10}$ this ratio was *ca.* 50 ± 20 % showing that the amount of R phase formed in $C_{16}/C_{16}EO_{10}$ samples was significantly higher than the one in $C_{16}/C_{16}SorbEO_{20}$.

In the $C_{16}/C_{16}EO_{10}$ samples, both R and C peaks co-existed in a wide temperature range down to 8.5-9.5°C where the peak for rotator phase disappeared completely in the sample with the smaller drops, **Figure 8**. Interestingly, in the sample containing bigger drops (initial diameter 33.5 µm), we also observed significant decrease of the rotator phase peak intensity at similar temperatures, but instead of disappearing, the peak shifted by ≈ 0.04 nm$^{-1}$ to $q ≈ 2.86$ nm$^{-1}$ and remained at this position with the same intensity down to 5°C, **Figure 7b**. Lower temperatures were not tested due to the limitation of the SAXS equipment.

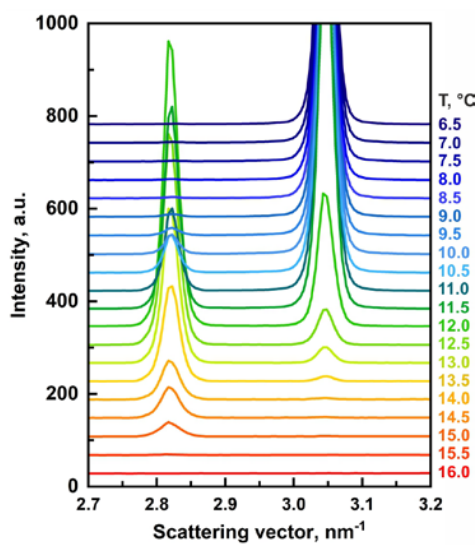

**Figure 8. SAXS curves obtained upon cooling of 15 µm $C_{16}$ drops dispersed in $C_{16}EO_{10}$ solution.** Cooling rate: 0.5°C/min. The R peak appears at 15.5°C and the C peak at ≈ 14.8°C in this sample (see also **Supplementary Figure S5**).

To understand better the obtained SAXS results, we performed additional optical observations with this emulsion (33.5 µm drops), lowering the temperature further down after the initial bright colors appeared (where we usually finished our previous experiments). We observed that most of the drops in this sample (> 80%) undergo process of plastification before their crystallization into triclinic crystalline phase. The R to C phase transition was observed in the microscopy experiments as the particle color changed abruptly and the particles slightly shrunk in size. These observations are illustrated with the images shown in **Figure 7c**, taken at 15°C and 13°C where the particles are still in rotator phase and their change upon further cooling to 10°C when the R→C phase transition has already occurred. The process is demonstrated also in

**Supplementary Movie 1**. Similar process of plastification followed by crystallization at lower temperatures was observed with the 15 μm drops as well, see **Supplementary Figure S6**.

To check whether the properties of the rotator phase formed in these drops are similar to those of the rotator phase observed in the limited fraction of hexadecane particles, stabilized by $C_{16}SorbEO_{20}$, we performed additional experiments with limited cooling, similar to those described in the previous section. The cooling without crystallization of the drops was realized using the following temperature protocol: the emulsion was cooled at 0.5°C/min constant rate from 23°C down to *ca.* 15.5°C where the rotator phase peak was visible and then the cooling was stopped. Afterwards, the temperature of the sample was kept constant for 5 min and then the sample was heated in a stepwise manner at every 5 min during which the temperature was maintained constant. Such experiments were performed with both the SAXS/WAXS instrument and the optical microscope. The obtained results are presented in **Supplementary Figure S7** and in **Figure 9**.

As expected, the SAXS/WAXS results obtained with $C_{16}EO_{10}$ surfactant were similar to those obtained with $C_{16}SorbEO_{20}$. The scattering profiles remained almost unchanged from the moment at which the temperature was held constant up to ≈ 15.5°C. Afterwards, the rotator phase peak began to decrease slowly its intensity, whereas the crystalline phase peak remained unchanged at this temperature, showing that no phase transition between the rotator and crystalline phases occurred in this temperature range. The rotator phase peak disappeared completely upon heating to 16.5-16.8°C. The small crystalline peak remained almost unchanged up to 17.6-17.7°C and disappeared quickly when the temperature was increased up to 18.1°C.

The optical observations with this system, however, revealed slightly different behavior than that observed with $C_{16}SorbEO_{20}$, see **Figure 9**. Instead of bulk melting of the particles upon temperature increase, the 33.5 μm drops stabilized by $C_{16}EO_{10}$ surfactant exhibited an initial separation (dewetting) of a liquid hexadecane phase from the plastified solid particles when the temperature was increased from 15.5 to 16°C. However, the colors of the particles did not change significantly showing that the liquid hexadecane did not form due to melting of the plastified hexadecane (otherwise, the intensity of the colors would have decreased significantly). All these results, combined together, indicate that some fraction of the hexadecane remained in a liquid state throughout this experiment, being trapped inside a thick shell of rotator phase. Careful check of the microscopy images confirmed the latter assumption – we could see a pocket of liquid hexadecane encapsulated inside the thick crust of plastic phase, see **Supplementary Figure S8**.

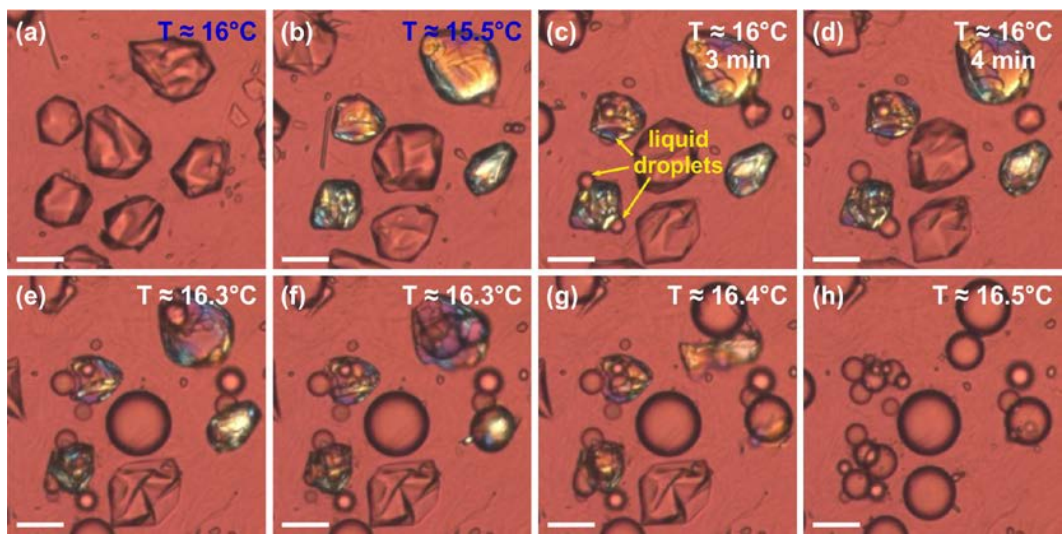

**Figure 9. Microscopy images of $C_{16}$ drops in $C_{16}EO_{10}$ at different temperatures.** The experiment is performed in the following way: the sample is cooled from 23°C to 15.3°C at 0.5°C/min rate, after which the cooling is stopped (images a and b). Then, the temperature is increased to 16°C and is kept constant for 5 min; afterwards the temperature is increased up to 16.5°C where all deformations disappear and the rotator phase melts. Scale bars, 20 μm.

The presence of liquid hexadecane inside the capsule of rotator phase could explain also the observed significant particle shrinkage observed upon cooling of these particles to lower temperatures, when the rotator phase transforms into a crystal phase, see **Figure 7c** and **Supplementary Movie 1**. As the rotator phase has much closer characteristics to the crystalline phase than the liquid phase, if the volume of the plastified particles was completely occupied by R phase, the shrinkage would have been much smaller than the one observed experimentally.

Upon further heating, we observed a complete melting of the shells of rotator phase at temperature ≈ 16.6 ± 0.2°C, in agreement with the results obtained with $C_{16}/C_{16}SorbEO_{20}$ emulsion. Again, the non-spherical drops observed at lower temperatures acquired spherical shape upon heating to this temperature, showing that the phase promoting the surface area increase had disappeared.

These X-ray scattering results explain the obtained DSC curve as well. The initial high peak observed at $T \approx 15°C$ is caused by the formation of rotator phase shells in most of the drops, whereas the continuous enthalpy gain observed between 14.5 and 11.5°C is caused by the freezing of the liquid particle interior and transformation of the plastic shells (composed of molecules arranged in rotator phase) into crystal triclinic phase. As already mentioned, the peaks at lower temperatures are caused by the freezing of the smaller droplets formed as a result of the breakage of the initial drops upon cooling. Indeed, if the entire particle volume was structured in R phase initially, then the observed peaks at lower temperatures should have been much smaller, because most of the enthalpy (65 to 75 %) in the full liquid-to-crystal transition is released during the L-to-R transition, whereas 25 to 35% of the enthalpy only is released in the subsequent R-to-C transition.[9,10,13]

Finally, we note another interesting observation in the SAXS spectra obtained at constant cooling. When the main part of the rotator phase peak ($q \approx 2.820 \pm 0.006$ nm$^{-1}$) disappears upon cooling down to 9°C, a small but well visible rotator phase peak remains at $q \approx 2.847 \pm 0.013$ nm$^{-1}$, **Figure 7b**. This peak was observed in all 3 independent experiments, performed with the 33.5 µm drops, whereas no such peak was observed in the experiments with the smaller drops. The difference in the peak maxima of *ca.* 0.027 nm$^{-1}$ corresponds to $\approx 0.02$ nm difference in the layer spacing. This difference is smaller even than the length of a single C-C or C-H bond which are in the range of 0.1 to 0.15 nm.[43] The rotator phase with $d \approx 2.207$ nm ($q \approx 2.85$ nm$^{-1}$) melted upon heating at $\approx 15$°C, **Figure 7d**. The significant difference between the melting temperatures of the two R phases, one melting at $\approx 15$°C and the other at $\approx 16.6$°C, indicate that these are two different rotator phases. Most probably, the peak shift is related to tilting with respect to the layer normal and the respective tilt angle can be calculated to be [$\pi/2$ - arcsin(2.207/2.228)] $\approx 8$°. Alternatively, the change in the spacing could be due to tighter packing of the molecules in the rotator phase, without any significant restructuring of the molecular positions, but then one could expect this process to be reversible and melting of this phase to happen at the same temperature, 16.6°C.

From the experiments with the system $C_{16}$ drops in $C_{16}EO_{10}$ solution we conclude that the initial thickness of the plastic rotator phase of 2-3 layers (if homogeneous thickness of the rotator phase is assumed as calculated by the DSC measurements[8]) is too small to induce sufficiently intensive diffraction peaks in this type of SAXS/WAXS measurements. On the other hand, we observed for the first time formation of micrometer thick shells of rotator phase around a liquid hexadecane drop interior. Upon further cooling, a small fraction of this rotator phase transformed into another rotator phase with smaller interlayer spacing, whereas the remaining bigger fraction underwent R-to-C transition. The two rotator phases melted at different temperatures, the phase with thickness 2.228 nm (q ≈ 2.82 nm$^{-1}$) had melting temperature of ≈ 16.6°C, whereas the phase with thickness 2.207 nm (q ≈ 2.85 nm$^{-1}$) melted at ≈ 15°C, see **Figure 10** for schematic illustration of these phases. These results indicate also that the surfactant molecules, inducing the rotator phase formation, and the size of the micro-confinement (the drop size in our experiments) affect strongly the characteristics of the formed R phase.

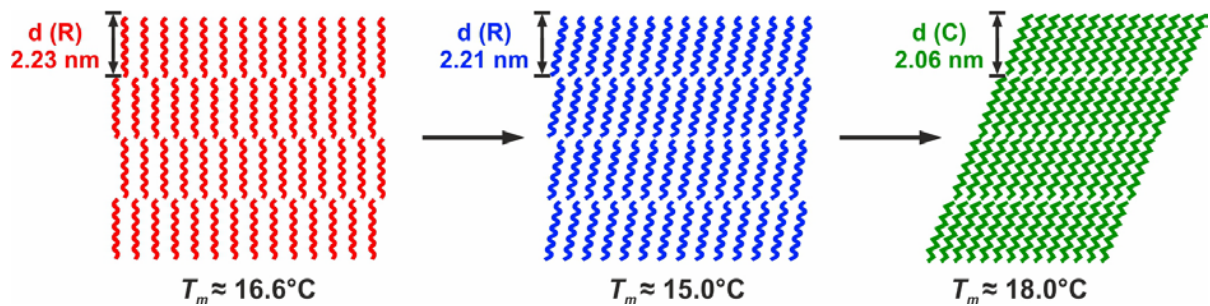

**Figure 10.** **Schematic presentation of the three phases which were observed in the current study with their spacing, melting temperatures and expected molecular packing.** The melting temperatures are determined upon heating. The two different rotator phase are observed only upon cooling (metastable character) when appropriate surface active compounds are used. Their melting is observed only if the cooling is stopped prior to transformation into more ordered phase.

**CONCLUSIONS**

The current study explores the structure of phases formed in micrometer-sized emulsion droplets able to morph through various non-spherical shapes upon cooling and prior to their freezing. In our previous studies[1-3,8] we suggest that these transformations are enabled by an enthalpy gain due to arrangement of part of the molecules situated next to the surface into the plastic rotator phases, however structural data was not available until now. Note that by definition the interfacial water-oil area increase is prohibited from the capillary pressure law which tries to minimize the energy of the systems.[43]

By using combination of SAXS/WAXS experiments with microscope observations and DSC measurements, in the current study we showed that rotator phase indeed form in micrometer-sized drops when the surfactant adsorption layers are able to freeze before the oil contained in the drops. Consequently, occurrence of non-spherical fluid shapes before the complete drop crystallization is observed for these systems. However, the initial freezing of the surfactant adsorption monolayer cannot be detected in the used experiments with emulsion drops placed in glass capillaries (it might be possible in other types of synchrotron experiments, e.g. with flat surface monolayers[17]). Thicker molecular multilayers should be formed to observe the appearance of rotator and/or crystalline peaks in the cumulative signal obtained with bulk emulsion samples.

The obtained results emphasize the importance of the type of the template molecules which induce the rotator phase formation. Furthermore, combining their role with the confinement effect known to increase the rotator phases stability[13,18-24,27-31], we show that hexadecane which exhibits only a very short-living transient rotator phase in bulk[12], may structure in long-living rotator phase in emulsion droplets. Formation of bulk rotator phases structuring the whole particle volume (plastification) was demonstrated for limited number of particles (0.2-0.3%) in the system $C_{16}$ drops in $C_{16}SorbEO_{20}$, whereas the formation of thick shells of rotator phase, surrounding liquid hexadecane interior, was observed in most (> 80%) of the $C_{16}$ drops dispersed in $C_{16}EO_{10}$ surfactant solution. The respective R phase was found to be thermodynamically metastable, because it was observed in a certain temperature interval upon cooling, whereas it did not form from the crystalline phase upon heating.

A characteristic layer spacing of ≈ 2.23 nm was determined for the surface rotator phase in the systems with self-shaping fluid droplets. This value is in good agreement with the previously reported results for the transient rotator phase formed in bulk hexadecane.[12] Furthermore, we determined for the first time the melting temperature of this thermodynamically metastable rotator phase to be, $T_{mR}$ ≈ 16.6 ± 0.3°C.

A second rotator phase with layer spacing of 2.21 nm, most probably containing tilted molecules with respect to the plane of the layers, was observed upon cooling of 33.5 μm $C_{16}$ drops dispersed in $C_{16}EO_{10}$ surfactant solution. This phase melted at significantly lower temperatures of $T_{mR}$ ≈ 15 ± 0.2°C.

The methodology combining SAXS/WAXS experiments with microscope observations and DSC measurements, opens almost unexplored opportunities for detecting and investigating the properties of rotator phases in micrometer-sized emulsion drops.[29] The additional phase characterization experiments used for the first time in the current study, may be applied for investigation of the effects of oil type, surfactants, drop size, electrolytes, temperature protocol and many other variables. The results obtained in the current study show that all these factors could change the formed polymorphic phases in unexpected and still poorly understood way. One can expand this approach also to investigate the kinetic aspects of the nucleation of rotator and crystal phases in micro-confinement. New questions related to the possible interplay between the surface and bulk nucleation in these systems can be formulated and answered. Therefore, we consider our current study also as an important methodological step towards the deeper structural investigation of these fascinating emulsion systems, promising the discovery of new phenomena triggered by the presence of phases with intermediate characteristics between the crystalline and liquid.

**Acknowledgements:**

The study received partial funding from the Bulgarian Ministry of Education and Science, under the National Research Program "VIHREN", project ROTA-Active (no. KP-06-DV-4/16.12.2019). Z.V. acknowledge the support from the Operational Program "Science and Education for Smart Growth", Bulgaria, grant number BG05M2OP001-1.002-0012. The research leading to these results has been supported by the project CALIPSOplus under Grant Agreement 730872 from the EU Framework Programme for Research and Innovation HORIZON 2020, under the execution of project 20202101 with proposer D.C. Preliminary SAXS experiments were performed in 2018 (Elettra Synchrotron, proposal 20182062; Soleil Synchrotron, proposal 20180355, proposer Dr. Sam Richardson) by D.C., D.G., Z.V. (Sofia University) and Dr. Stoyan Smoukov, Ms. Artie Julian Herbert and Mr. Merhdad Isfandbod (QMUL) – no results from these preliminary experiments are included in the current paper. The authors thank Dr. Barbara Sartori and Dr. Heinz Amenitsch (Elettra Sincrotron, Trieste, Italy) for their valuable help during the SAXS/WAXS measurements included in the paper.

**Contributions:**

D.C. – conceptualization; methodology; investigation; validation; formal analysis; visualization; writing – original draft, review and editing; project administration; funding acquisition

D.G. – investigation; validation; visualization; formal analysis

Z.V. – investigation; visualization

S.Ts. – investigation

K.T. – investigation

S.Tc. – conceptualization; methodology; supervision; writing – review and editing; funding acquisition

N.D. – conceptualization; methodology; supervision; writing – review and editing; funding acquisition

*Paper accepted for publication in Journal of Colloid and Interface Science, doi: 10.1016/j.jcis.2021.06.122*33. Denkov, N.; Tcholakova, S.; Cholakova, D. Surface phase transitions in foams and emulsions. *Curr. Opin. Coll. Interf. Sci.*, **2019**, *44*, 32-47. doi: 10.1016/j.cocis.2019.09.005

34. Haas, P.; Goldstein, R.; Smoukov, S. K.; Cholakova, D.; Denkov, N. Theory of shape-shifting droplets. *Phys. Rev. Lett.*, **2017**, *118*, 088001. doi: 10.1103/PhysRevLett.118.088001

35. Haas, P.; Cholakova, D.; Denkov, N.; Goldstein, R.; Smoukov, S. Shape-shifting polyhedral droplets. *Phys. Rev. Research.* **2019**, *1*, 023017. doi: 10.1103/PhysRevResearch.1.023017

36. Garcia-Aguilar, I.; Fonda, P.; Sloutskin, E.; Giomi, L. Faceting and flattening of emulsion droplets: A mechanical model. *Phys. Rev. Lett.* **2021**, *126*, 038001. doi: 10.1103/PhysRevLett.126.038001

37. Nakashima, T.; Shimizu, M.; Kukizaki, M. Membrane emulsification by microporous glass. *Key Eng. Mater.* **1992**, *61-62*, 513-516. doi: 10.4028/www.scientific.net/KEM.61-62.513

38. Joscelyne, S. M.; Tragardh, G. Membrane emulsification – A literature review. *J. Membrane Sci.* **2000**, *169*, 107-117. doi: 10.1016/S0376-7388(99)00334-8

39. Newton, R. H.; Haffegee, J. P.; Ho, M. H. Polarized light microscopy of weakly birefringent biological specimens. *J. Microsc.* **1995**, *180*, 127-130. doi: 10.1111/j.1365-2818.1995.tb03667.x

40. Holmberg, K. Handbook of applied surface and colloid chemistry. Vol. 2 Ch. 16 Identification of lyotropic liquid crystalline mesophases. *John Wiley & Sons* **2001**, 299-332. ISBN: 978-0-471-49083-8

41. Petrovic, M.; Barcelo, D. Analysis of ethoxylated nonionic surfactants and their metabolites by liquid chromatography/atmospheric pressure ionization mass spectrometry. *J. Mass Spectrom.* **2001**, *36*, 1173-1185. doi: 10.1002/jms.234

42. Sparham, C. J.; Bromilow, I. D.; Dean, J. R. SPE/LC/ESI/MS with phthalic anhydride derivatisation for the determination of alcohol ethoxylated surfactants in sewage influent
30

# Supplementary materials

# Rotator phases in hexadecane emulsion drops revealed by X-ray synchrotron techniques

**Diana Cholakova, Desislava Glushkova, Zhulieta Valkova**
**Sonya Tsibranska-Gyoreva, Krastina Tsvetkova**
**Slavka Tcholakova, Nikolai Denkov***

*Department of Chemical and Pharmaceutical Engineering*
*Faculty of Chemistry and Pharmacy, Sofia University,*
*1 James Bourchier Avenue, 1164 Sofia, Bulgaria*

*Corresponding author:
Prof. Nikolai Denkov
E-mail: nd@lcpe.uni-sofia.bg
Tel: +359 2 8161639
Fax: +359 2 9625643

**Supplementary Movie 1.** Microscopy observation of hexadecane drops, dispersed in 0.75 wt. % $C_{16}EO_{10}$ surfactant solution, upon cooling at a rate of 0.5°C/min. The initial drop shape deformations start at around 22-21°C for the main fraction of the drops, whereas the complete drop plastification (L to R phase transition of the drop interior) occurs at ≈ 15°C. The rotator-to-crystal phase transition occurs at lower temperatures (11.5 to 8°C) and it is seen as a change of the particle colors and noticeable particle shrinkage.

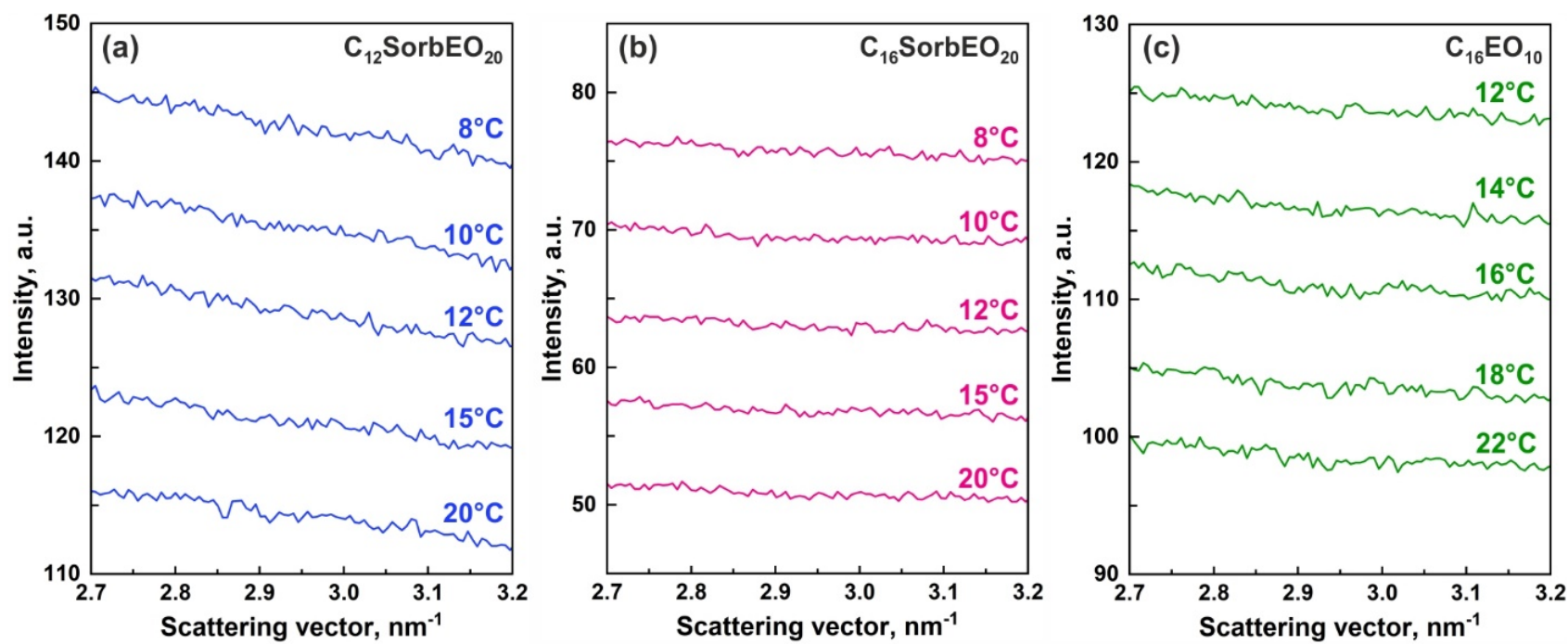

**Supplementary Figure S1.** SAXS spectra obtained upon cooling of various surfactant solutions at a rate of 0.5°C/min. **(a)** SAXS spectra of 1.5 wt. % $C_{12}SorbEO_{20}$ solution measured at different temperatures. No peaks are detected and the intensity of the scattered signal remains constant between 20°C and 8°C. **(b)** SAXS spectra of 1.5 wt. % $C_{16}SorbEO_{20}$ solution in the temperature range between 20°C and 8°C. **(b)** SAXS spectra of 0.75 wt. % $C_{16}EO_{10}$ solution in the temperature range between 22°C and 12°C. None of the solutions undergoes bulk structural changes detectable by SAXS in the relevant *q*-range in the temperature range of interest.

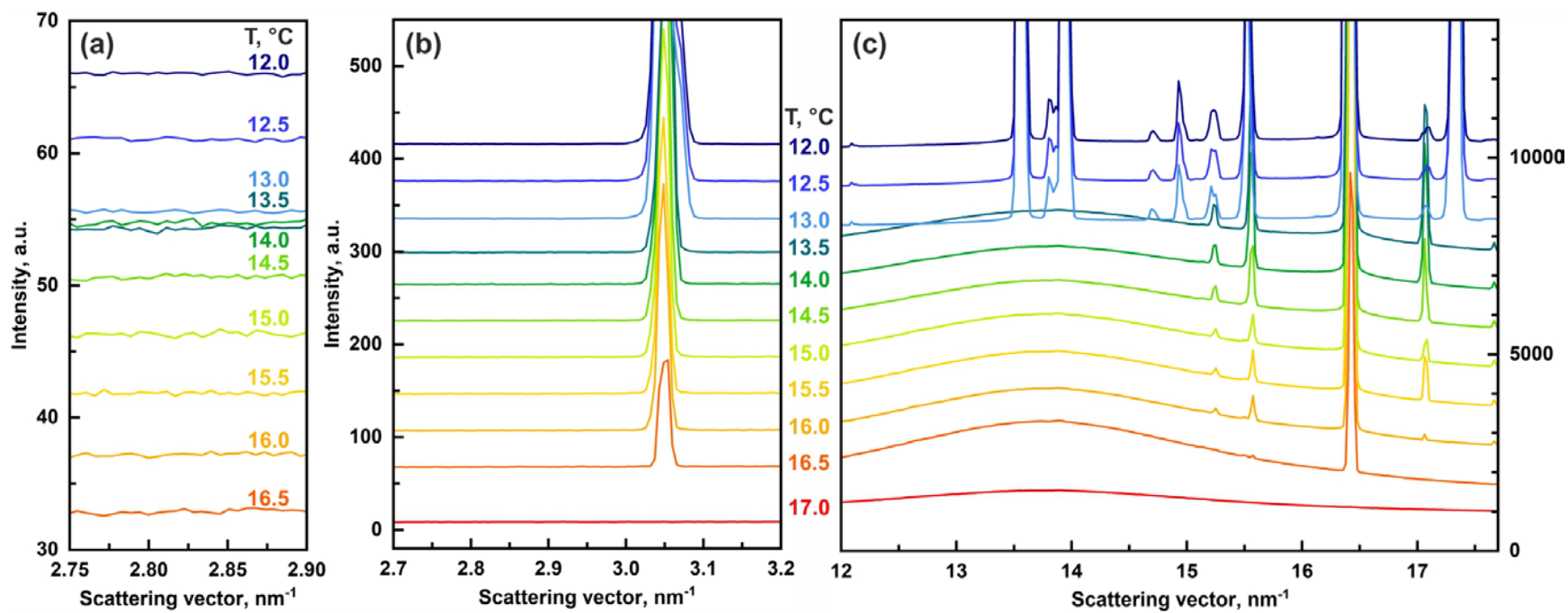

**Supplementary Figure S2. SAXS (a,b) and WAXS (c) spectra obtained upon cooling of bulk hexadecane at 0.5°C/min.** Formation of rotator phase is not observed in the sample of bulk alkane when using the temperature protocol and detection technique applied in the current study to the hexadecane-in-water emulsions. The peaks observed in the SAXS and WAXS spectra indicate the formation of the fully crystalline triclinic structure.

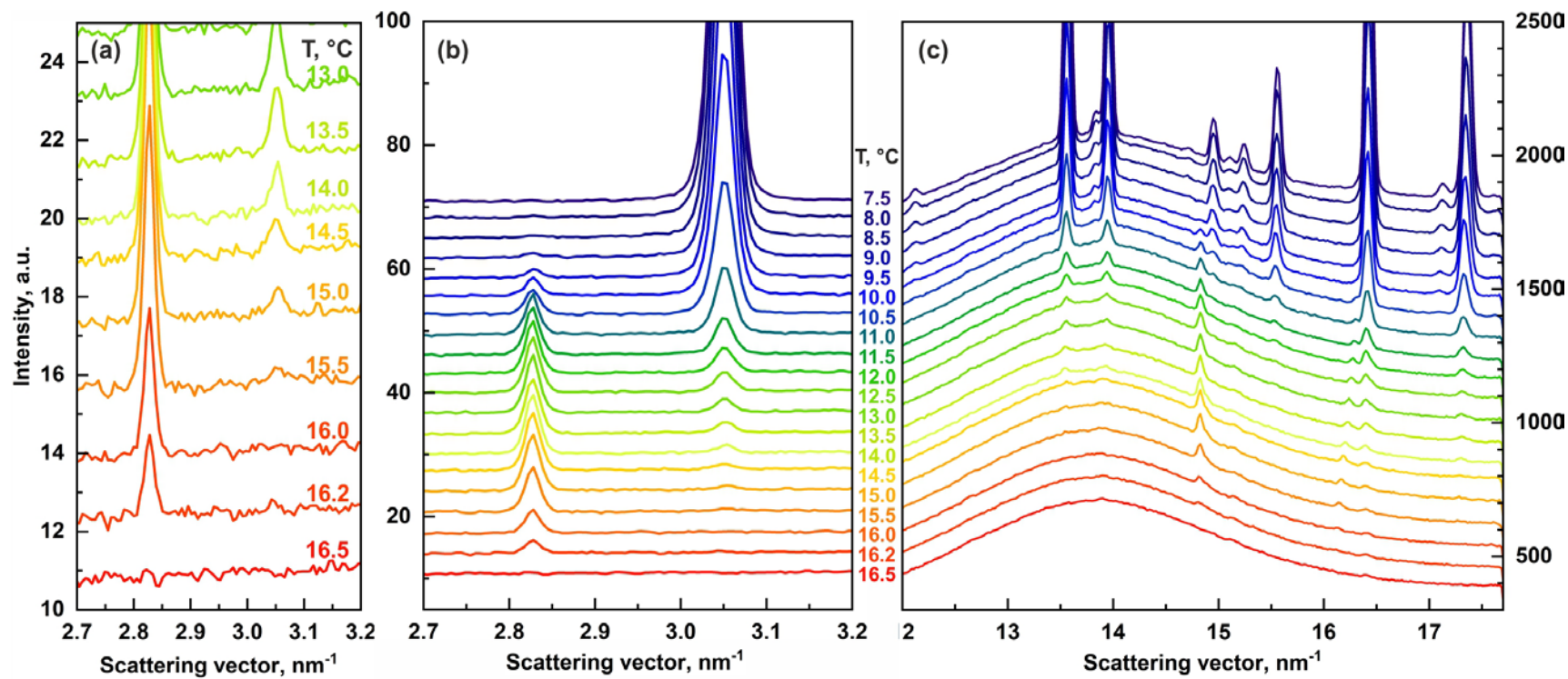

**Supplementary Figure S3.** SAXS (a,b) and WAXS (c) spectra obtained upon cooling of 12.1 μm hexadecane drops, dispersed in 1.5 wt. % $C_{16}SorbEO_{20}$ solution at a cooling rate of 0.5°C/min. The temperature for each curve on (b) and (c) is given between the two graphs. Note the different scales of the y-axes in these plots. The curves are shifted with respect to the *y*-axis for clarity.

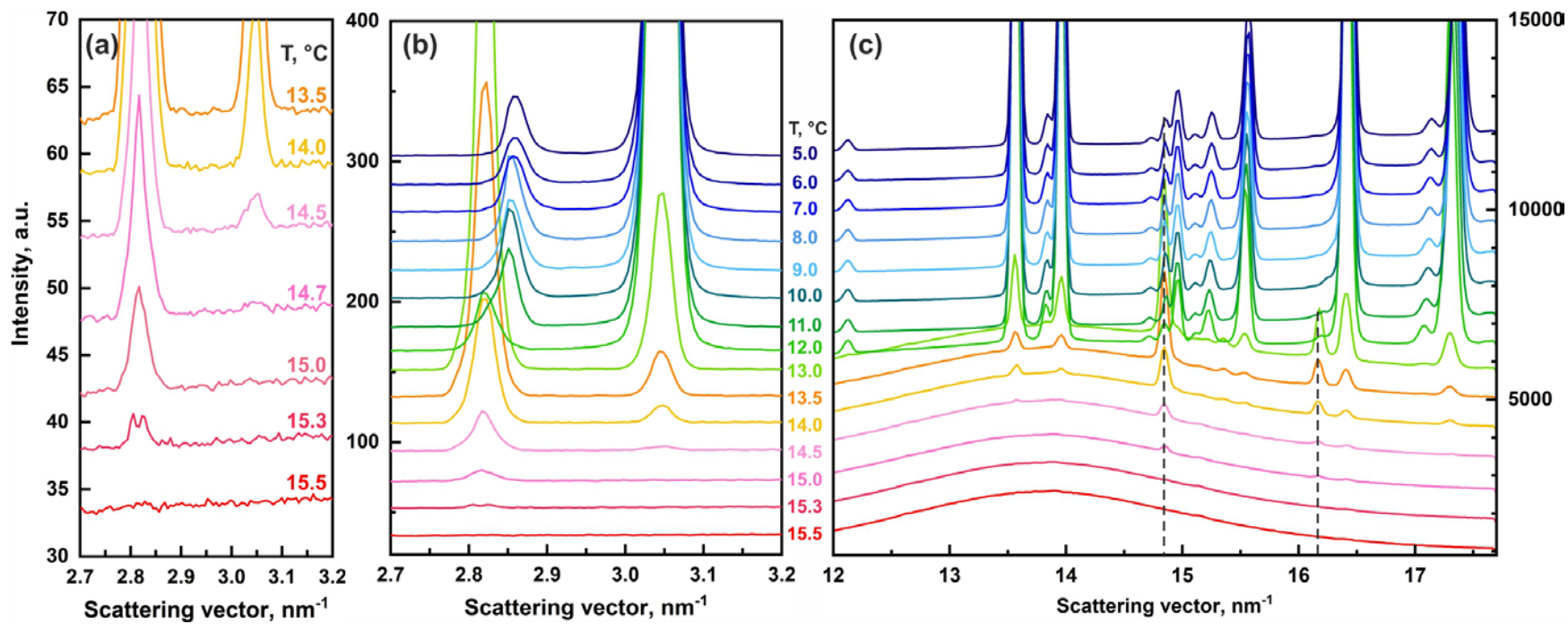

**Supplementary Figure S4.** SAXS (a,b) and WAXS (c) spectra obtained upon cooling of 33.5 μm hexadecane drops, dispersed in 0.75 wt. % $C_{16}EO_{10}$ solution, at a cooling rate of 0.5°C/min. Note that the peaks characterizing the rotator phase remain present in the spectra even at 5°C, as well as the change in the position of the peak coming from the rotator phase in (b) in the temperature range between 12 and 11°C.

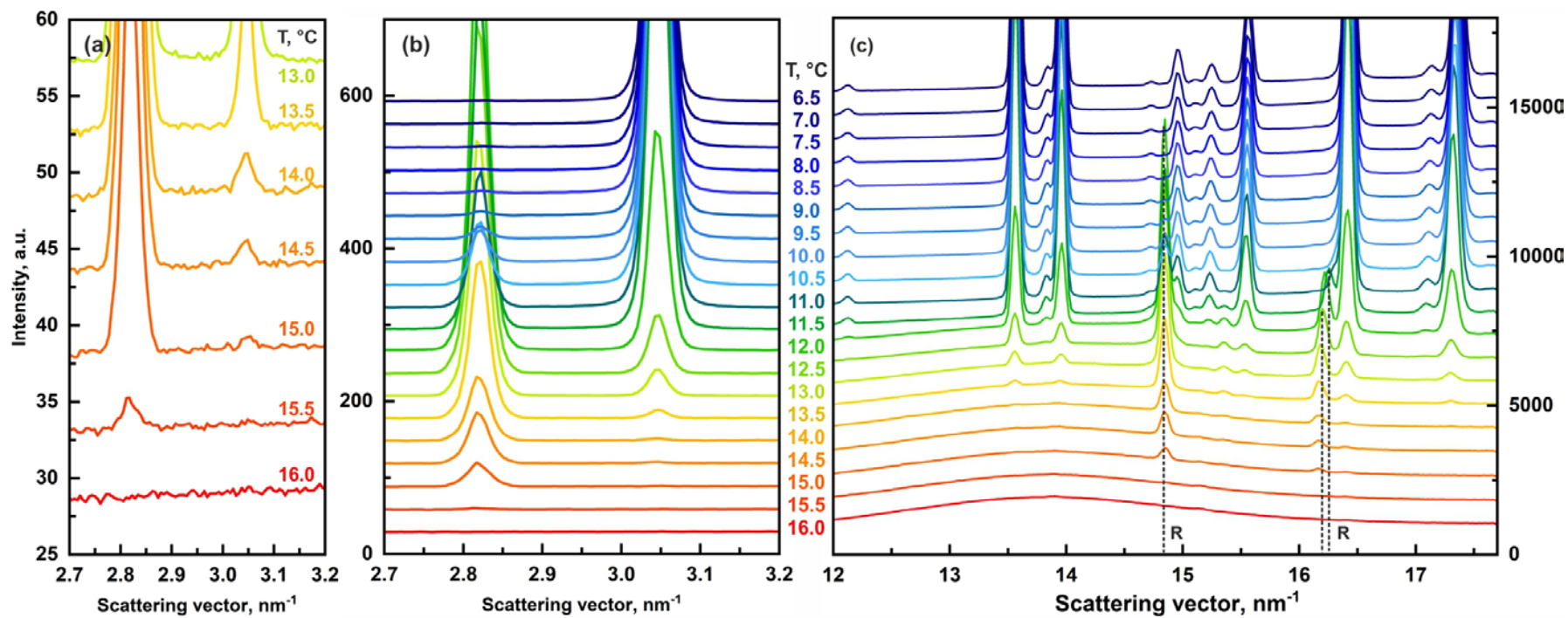

**Supplementary Figure S5.** SAXS (a,b) and WAXS (c) spectra obtained upon cooling of 15 μm hexadecane drops, dispersed in 0.75 wt. % $C_{16}EO_{10}$ solution, at a cooling rate of 0.5°C/min.

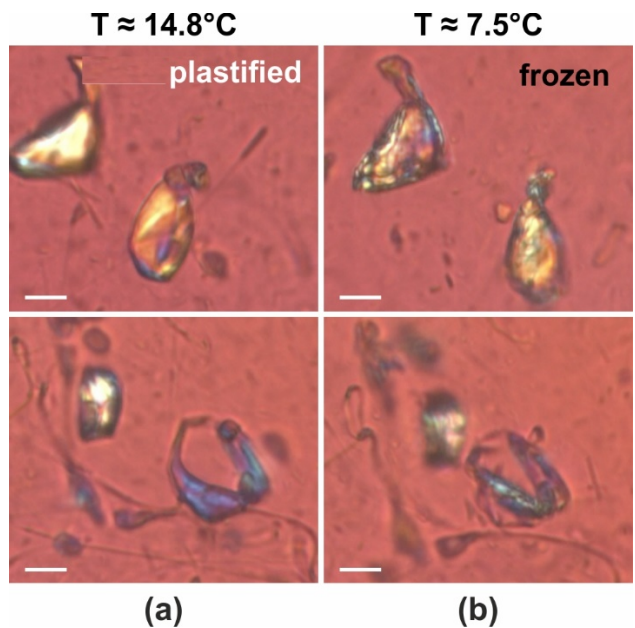

**Supplementary Figure S6. Microscopy images of plastified (a) and frozen (b) hexadecane particles.** The experiment is performed at 0.5°C/min cooling rate with 15 μm hexadecane drops, dispersed in 0.75 wt. % $C_{16}EO_{10}$ surfactant solution. Note the significant change in the colors of the particles after the R-to-C phase transition. Scale bars, 10 μm.

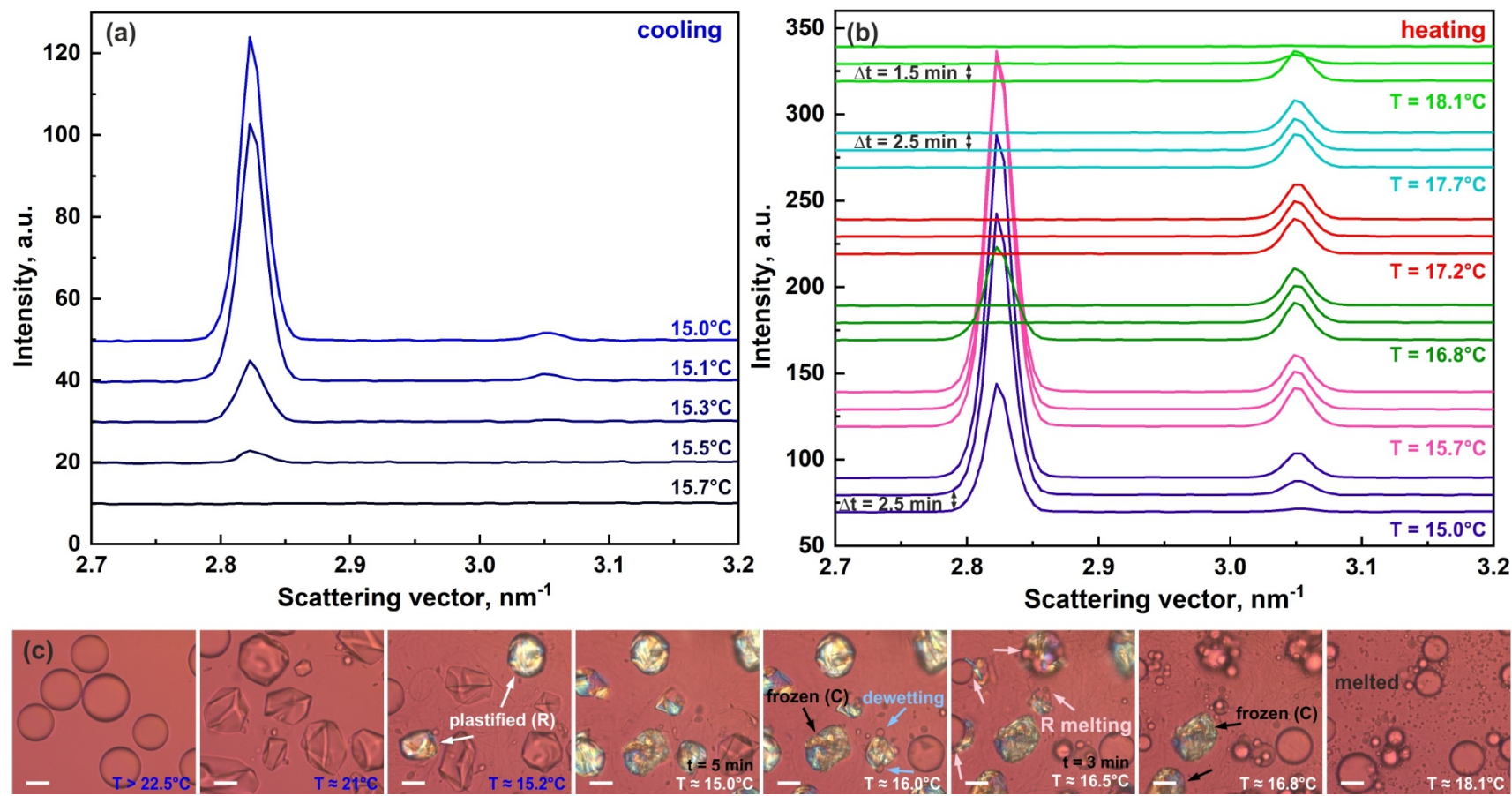

**Supplementary Figure S7.** **SAXS spectra (a,b) and microscopy images (c) of 33.5 μm $C_{16}$ drops dispersed in $C_{16}EO_{10}$ solution.** The SAXS experiment is performed using the following temperature protocol: **(a)** the sample is cooled from 23°C down to 15°C at 0.5°C/min cooling rate. **(b)** Afterwards, the temperature is kept constant for 5 min. Next, the temperature is increased in a stepwise manner to 15.7°C; 16.8°C; 17.2°C; 17.7°C and 18.1°C and is kept constant for 5 min at each temperature. The optical microscopy experiment is performed in the following way: the sample is cooled down from 23°C to 15°C; afterwards the cooling is stopped and the temperature is increased at 0.5°C steps – at each step, the temperature is kept constant for 5 min. The selected images show the most important processes occurring in the sample. Scale bars, 20 μm.

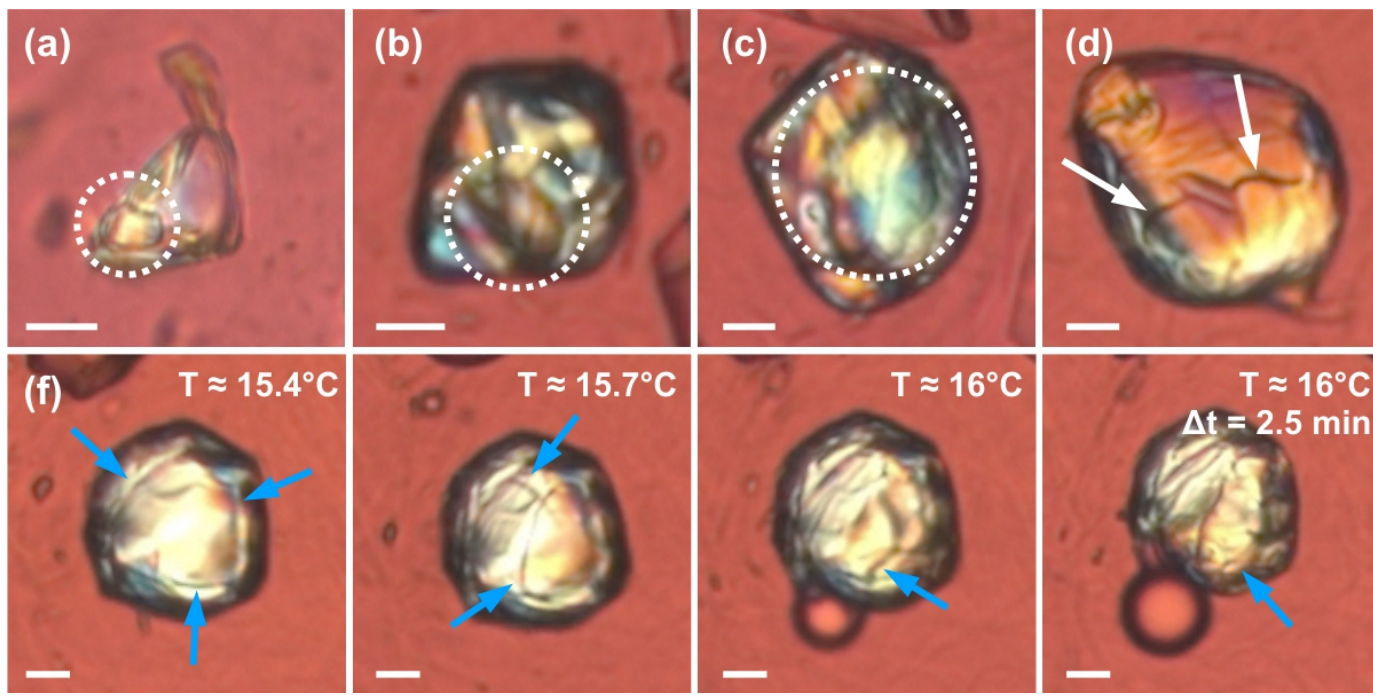

**Supplementary Figure S8.** **Microscopy images of plastified C$_{16}$ particles, dispersed in C$_{16}$EO$_{10}$ surfactant solution. (a-e)** Particles as observed after the plastification process. The white circles and arrows show the liquid component which remains trapped inside the thick crust of rotator phase (see the different contrast inside the particles). **(f)** Microscopy pictures of plastified particles, containing inner liquid hexadecane pocket, as observed immediately after the plastification (15.4°C) and after slight heating as indicated by the temperature labels on the pictures. Note that a dewetting process with a separation of oil drops occurred prior to the melting of the R phase. The particle colors did not changed upon this dewetting, showing that the inner pocket of oil had remained in a liquid throughout the entire experiment (even at 15.4°C). Scale bars, 10 μm.